\documentclass[11pt,twoside]{article}

\usepackage{asp2006}
\usepackage{epsf}
\usepackage{lscape}

\begin{document}
\title{Hydrogen Lyman emission through the solar cycle}
\author{Werner Curdt$^1$ and Hui Tian$^{1, 2}$}
\affil{\it $^1$MPI f\"ur Sonnensystemforschung, 37191 Katlenburg-Lindau, Germany\\
       $^2$School of Earth and Space Sciences, Peking University, China}

\begin{abstract}

We present observations and results of radiance and irradiance studies completed
by {\it SOHO}--SUMER during the past solar cycle. We find that the cycle variation in Ly$-\alpha$ irradiance
as observed by e.g. {\it UARS}--SOLSTICE can not be explained by quiet sun radiance data, and
conclude that the explanation must be related to differences in the Ly$-\alpha$
radiance of various solar features and changes in their fractional distribution
over the solar cycle. Consequently, we studied the emission of the hydrogen
Ly$-\alpha$ line in various solar features –- for the first time observed by SUMER
on disk in full resolution –- to investigate the imprint of the magnetic field on line profile and radiance
distribution. We also compare quasi-simultaneous Ly$-\alpha$ and Ly$-\beta$ line profiles.
Such high-resolution observations –- not hampered by geocoronal
absorption –- have never been completed before.

\end{abstract}

\section{Radiance and Irradiance}

The spectral solar irradiance varies with the solar cycle. The variation,
which is 0.1\,\% in visible light, increases significantly at shorter
wavelengths. It reaches 15\,\% at 1600~\AA, 50\,\% at 1216~\AA, a factor of 5
around 300~\AA, and two orders of magnitude in X-ray \citep[e.g.,][]{Frohlich09}.
The variation of 50\,\% is also seen in SUMER Ly$-\alpha$ data; during 'Sun as a star'
observations, the spectrometer slit was pointed far off-disk many
times through the solar cycle to let the spectrometer analyze the scattered light
\citep{Lemaire05}. These data are irradiance measurements, since the entire
disk contributes to the scattered light (although not at an equal share).

\begin{figure} 
  \plottwo{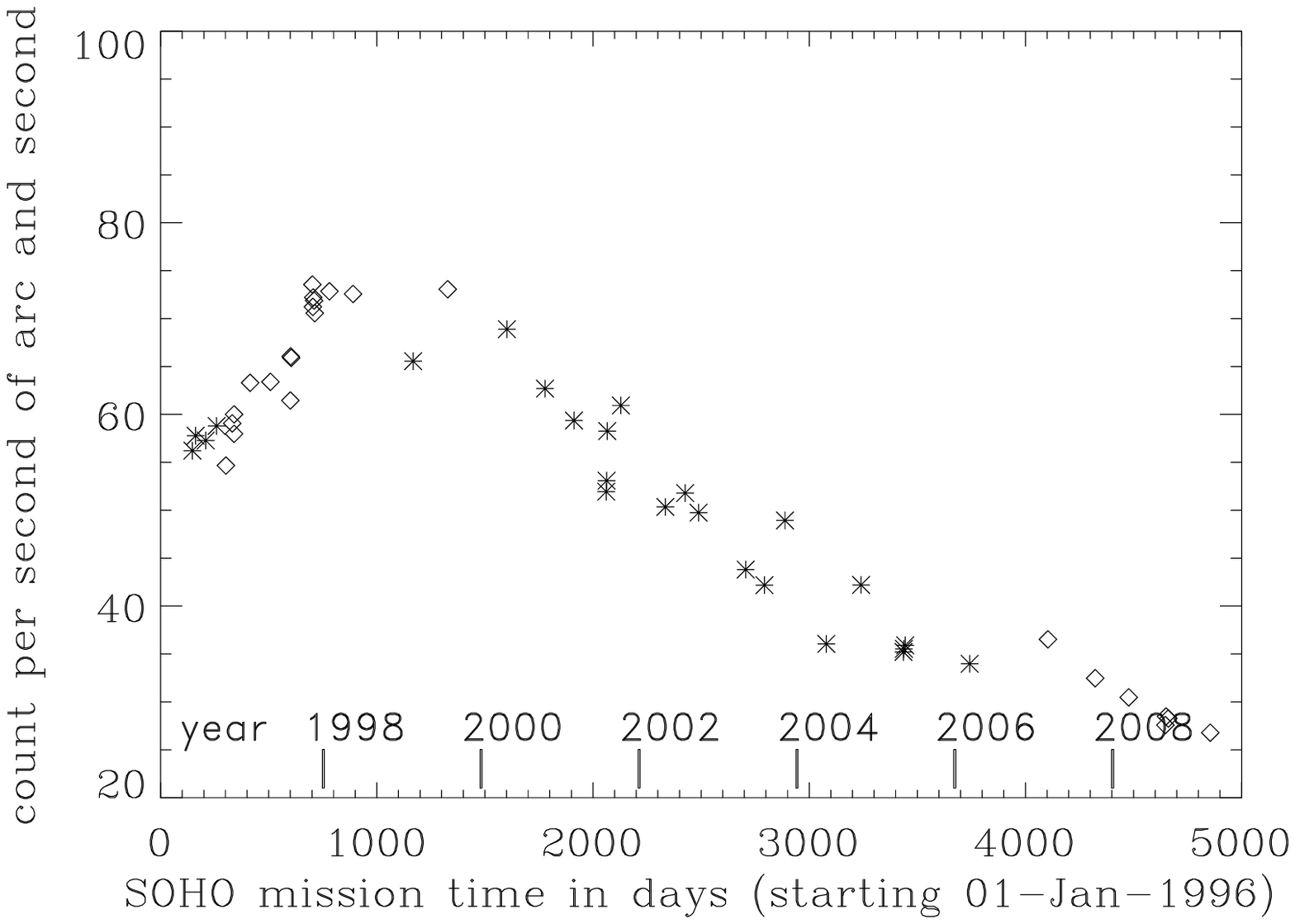}{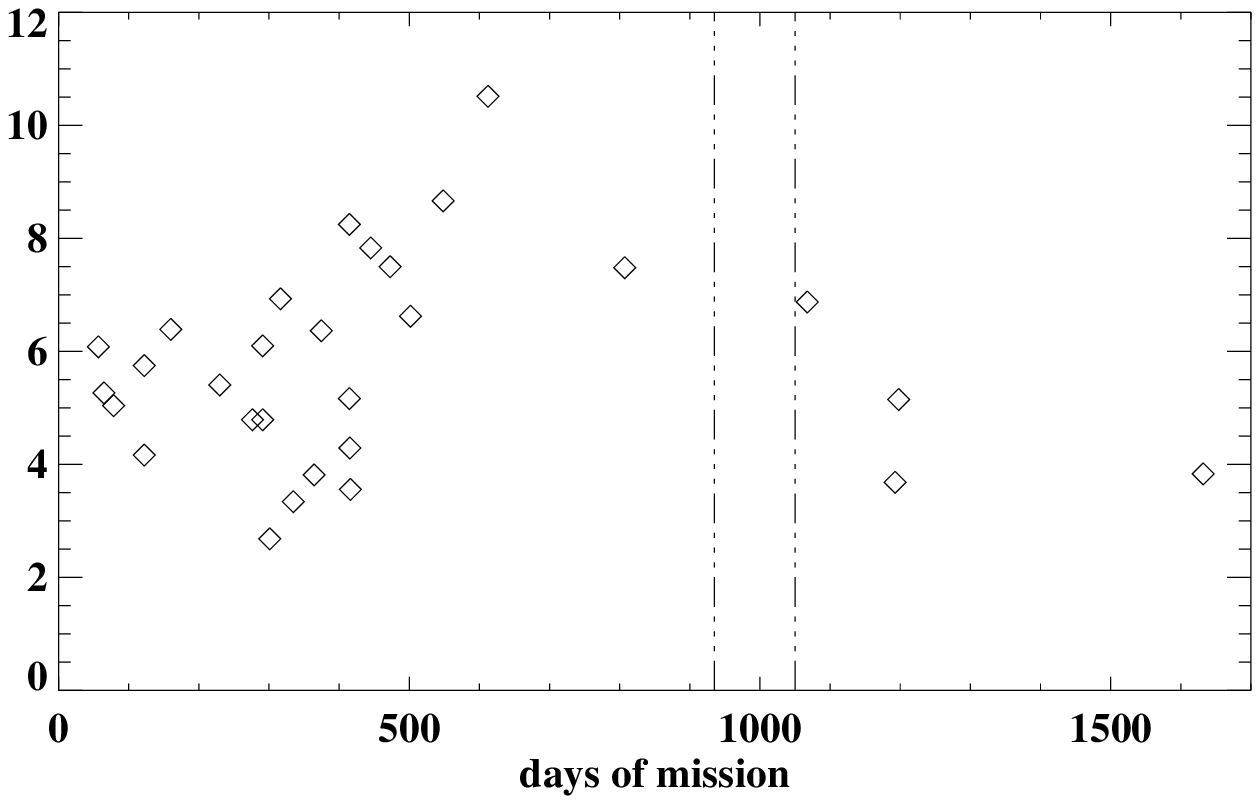}
  \caption {Left: Variation of the total scattered radiation observed at 800" above the
  limb in the Ly$-\alpha$ line \citep[courtesy: ][asterisk and diamond denote A or B detector]{Lemaire09}.
  The sensitivity loss of 40\,\%, which SUMER experienced during the SOHO accident is not
  taken into account.\newline
  Right: Radiance of the Lyman continuum in the quiet Sun during the rise of solar cycle 23
  \citep[cf.~for details,][]{Schuhle00}.}
\end{figure}

In order to track the responsivity of the SUMER instrument and to allow intercalibration
with other instruments, relatively large, standardized  rasters of quiet Sun
were completed until the year 2000 (so-called intercal JOPs), when it became
clear that -- with SUMER not doing rasters anymore -- it was difficult to obtain
data representative of the average quiet Sun. The results from
these observations were presented by \citet{Schuhle00}. If at all, only a
small uptrend is seen in Fig.~2 of that paper. We assume that some of
the rasters were contaminated by bright points or plage pixels, since during
solar maximum large enough patches of quiet Sun were scarce and conclude,
that the radiance of the average quiet Sun is more-or-less unchanged over the
cycle. The variation seen in irradiance data seems to be related to the fact,
that during maximum the percentage of quiet Sun regions goes down and quiet Sun
is replaced by brighter stuff.
It is therefore essential to study the variation of Ly$-\alpha$ radiances for
different solar disk features. This was, however, not possible with standard
observations, because Ly$-\alpha$ is much too bright and would saturate the detector.
In 2008 SUMER found a way to overcome this problem and started
unconventional observations, where 80\,\% of the aperture is vignetted by the
partially opened door. Some of the results found so far are described below.

\section{Ly$-\alpha$ and Ly$-\beta$ comparison}

Several of these rasters were completed at various $\mu$ angles in June, July
and September 2008, when the Sun was very quiet. In April 2009 also coronal hole data was
obtained. In some of the rasters, the wavelength mechanism was switched back
and forth between Ly$-\alpha$ and Ly$-\beta$, so that quasi-simultaneous data could be
recorded for those lines. The elapsed time between the exposure start of
consecutive frames was only 25.3~s, and we assume only minor temporal variations of the emitting
plasma; for observational details see \citet{Curdt08b, Tian09a, Tian09b}.
In the same exposure, we also recorded a
transition region line as velocity indicator. In the case of Ly$-\alpha$,
the Doppler flow in each pixel has been determined from the
$\lambda$\,1206\,Si\,{\sc{iii}} line; negative values stand for upflows. The
complex profile of the Ly$-\alpha$ line -- the central depression leaving red and blue peaks --
does not allow Doppler neasurements.
We sort the Ly$-\alpha$ profiles by the Doppler flow in each pixel
and define six equally spaced velocity bins. The profiles of each bin are
displayed in Fig.2, left panel.
The blue peak is always stronger than the red peak.
Both peaks and the central reversal are offset towards the red with increasing downflow,
and the asymmetry rises with increasing redshift.
Similarly, we display in Fig.2, right panel the profiles of Ly$-\beta$.
Although there is an obvious correspondence between asymmetry and downflows
for both lines, it is also clear that the asymmetries of Ly$-\alpha$ and Ly$-\beta$ are
reversed. This last, not obvious, result is only slightly reproduced by
Fontenla's \citeyear{2002} model with stronger downflow. The relationship between
asymmetry and downflows and the comparison with models would, thus,
indicate the presence of a persistent downflow like that in optically thin
transition region lines.

\begin{figure} 
  \plottwo{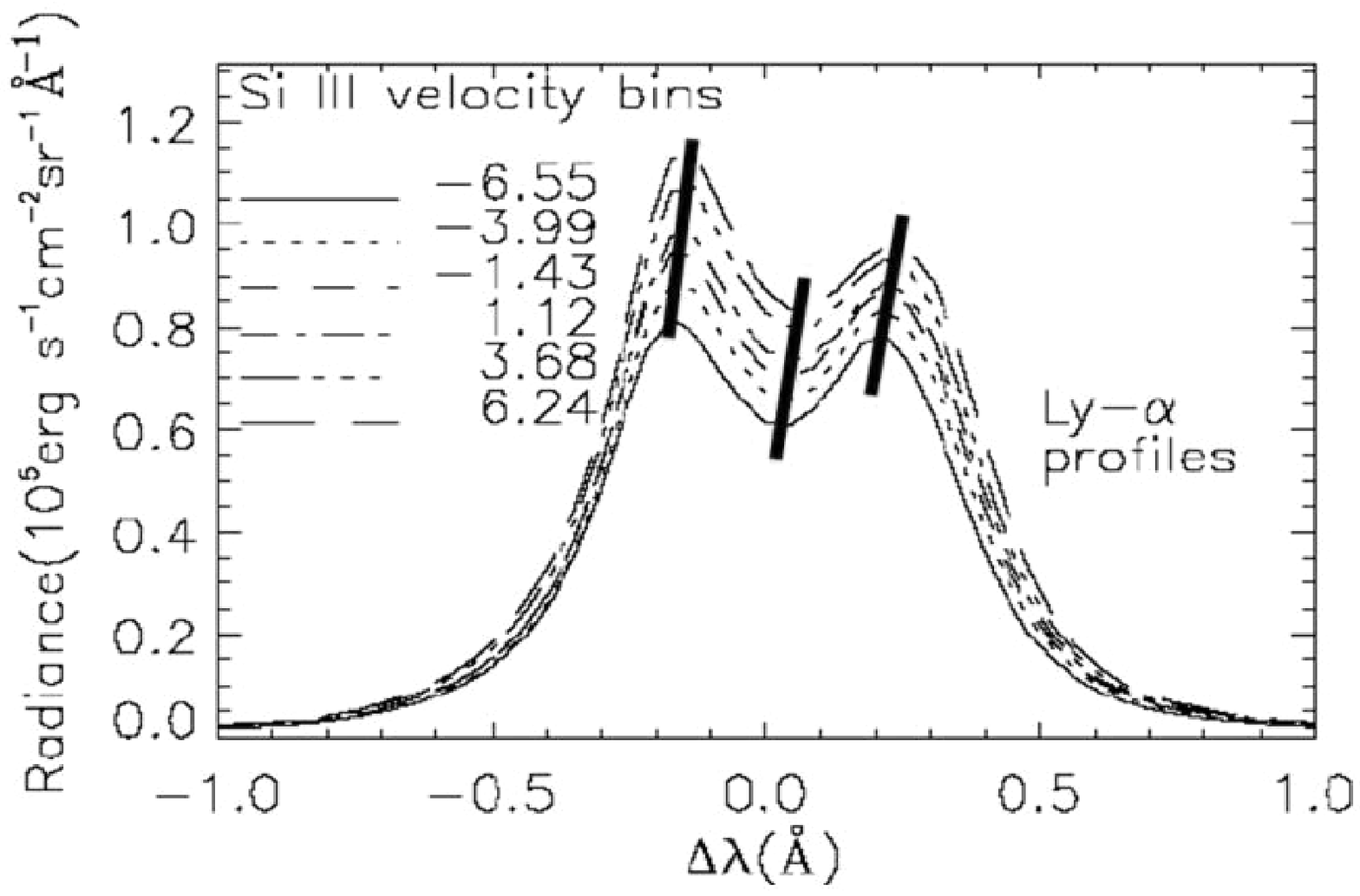}{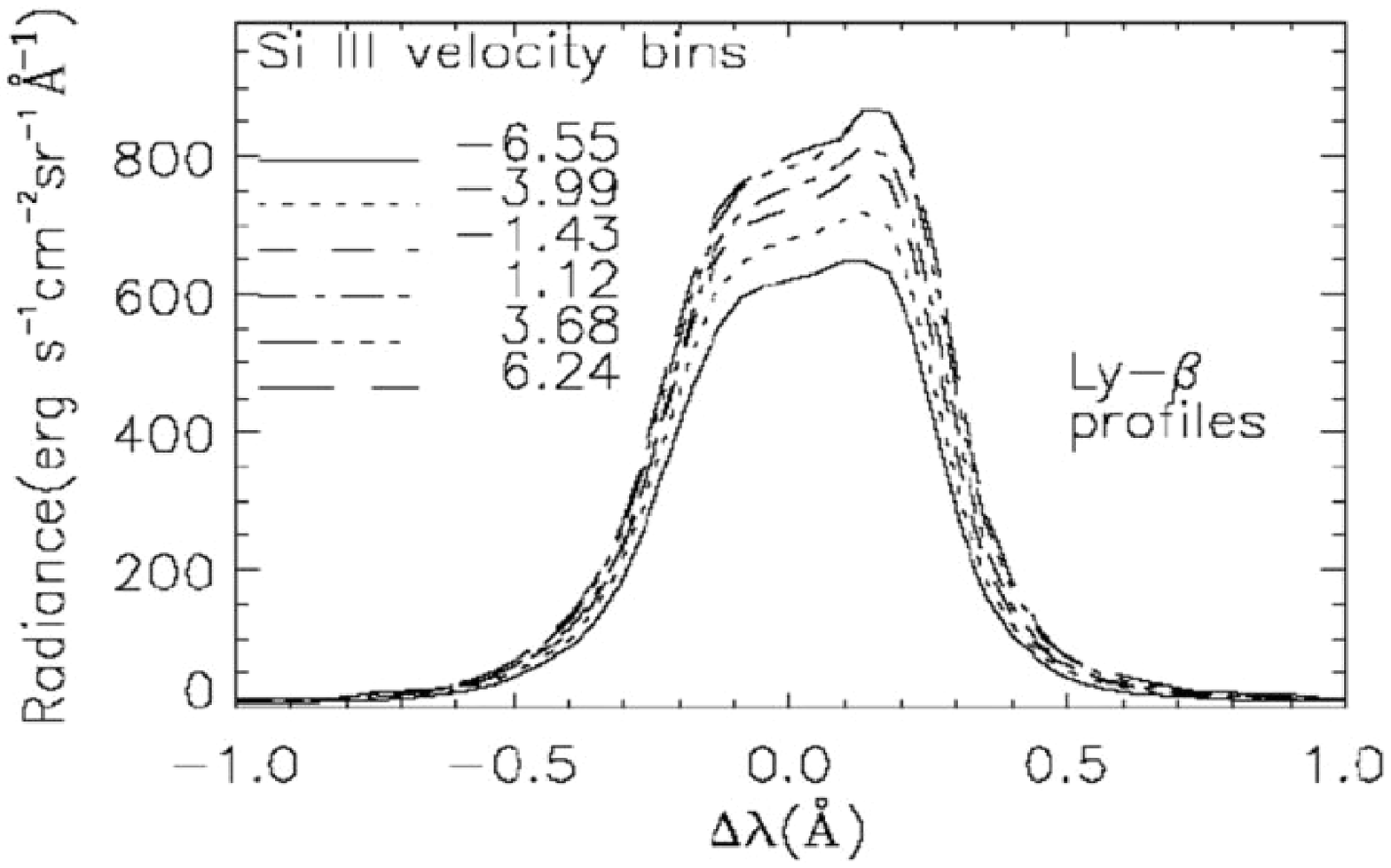}
  \caption {Comparison of quasi-simultaneous line profiles of Ly$-\alpha$ and Ly$-\beta$
  observed near disk center.  We show the profiles for six equally spaced velocity
  bins, which were defined by a pixel-by-pixel comparison of the
  $\lambda$ 1206 Si\,{\sc{iii}} line centroids with the rest wavelength.
  Negative values correspond to upflows, positive values to downflows.
  It is most obvious that the asymmetries in the Ly$-\alpha$ (left) and the Ly$-\beta$
  lines (right) are reversed, and there is a clear correspondence
  between asymmetry and downflows for both lines.}

\end{figure}

This surprising result is the combined effect of opacity and transition
region downflow and has a straightforward empirical explanation:
Both lines are optically thick, but the opacity in Ly$-\beta$
is much less than that in Ly$-\alpha$. In Ly$-\beta$, we can still see the redshifted
footpoints shining through, and its behavior is similar to that of a
transition region line, which is redshifted because of the (statistical)
redhift-radiance relationship \citep{IED08}. The opacity of Ly$-\alpha$ is so high,
that any directional information is lost through the partial redistribution
(PRD) process. In contrast to Ly$-\beta$ and all the higher Lyman lines, which are still dominated by
redshifted {\it emission} (which narrows the blue peak), Ly$-\alpha$ is dominated by
redshifted {\it absorption} (which suppresses the red peak).

We are convinced that this is the correct interpretation, and as a proof
we take our recent observations in a coronal hole.

In the coronal hole we found the largest peak separation in the Ly$-\alpha$ profile
(cf., Fig.~3) and the deepest self-revearsals in the Ly$-\beta$ profile, in both cases
indication of extremely high opacity \citep{Tian09b}. It is so high that now
-- similar to Ly$-\alpha$ -- also Ly$-\beta$ is observed with blue-peak asymmetry.
Xia \citeyear{Xia03} already found in his PhD work, that in coronal holes the fraction of profiles
with blue-peak dominance is increased, but his results did not allow a consistent interpretation.

\begin{figure} 
  \plotone{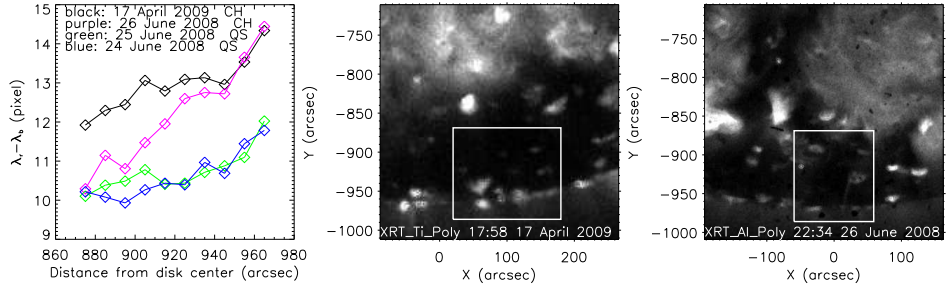}
  \caption {The separation of the red and the blue peak of Ly$-\alpha$ in the polar
  region. Approaching the pole, a signature is seen in coronal hole spectra:
  The peak separation increases significantly, indication of higher
  opacity \citep[cf. for details,][]{Tian09b}.}

\end{figure}

The photon ratio of both Lyman lines varies between 132 and 255
without a clear correspondence to the network pattern.
The scatter plot of Ly$-\alpha$ versus Ly$-\beta$ photons (not displayed here) shows a linear relationship with
a slope  of 188 $\pm$ 1.  This value is much higher than those reported in
literature so far. It can also be used to constrain the intercalibration
of both the Ly$-\alpha$ and the O\,{\sc{vi}} channel of the UVCS instrument \citep{Kohl97}.

\section{The Ly$-\alpha$ profile and the magnetic field}

Magnetograms in grey, overlaid with Ly$-\alpha$ brightness contours (white) nicely
outline the network as shown in Fig.4. Before starting a more detailed analysis
of the profiles we applied a smooth-3 operator to the profiles in each
spatial pixel. This helped to reduce the noise and to
improve the statistics. We have sorted the pixels by the depth of their
self-revearsal, $D_i$, and defined four bins separated by the quartiles,
$D_{q1}$, $D_{q2}$, and $D_{q3}$. In Fig.4, pixels in the extreme bins are marked by
white triangles\footnote{A better readable coloured version of this figure is available
under\\ $http://www.mps.mpg.de/homes/curdt/Lyman.pdf$ .}, patches with deep
revearsal on the right side and patches with a flat profile on the left side.
It is obvious, that the deeply reversed profiles cluster in the cell interior
and the more flat profiles in the network.
This suggests a lower opacity in the network funnels made of high-lying loops
as compared to the low-lying loops in the cell interior. As mentioned before
in coronal hole spectra signatures of even higher opacity were found. Higher
opacity is equivalent to more neutral hydrogen. This could mean that the network
funnels in coronal holes do not open as wide as they do in the quiet Sun.
But also a temperature decrease would affect the degree of ionization and the
population of the ground level of neutral hydrogen.

\begin{figure} 
  \plotone{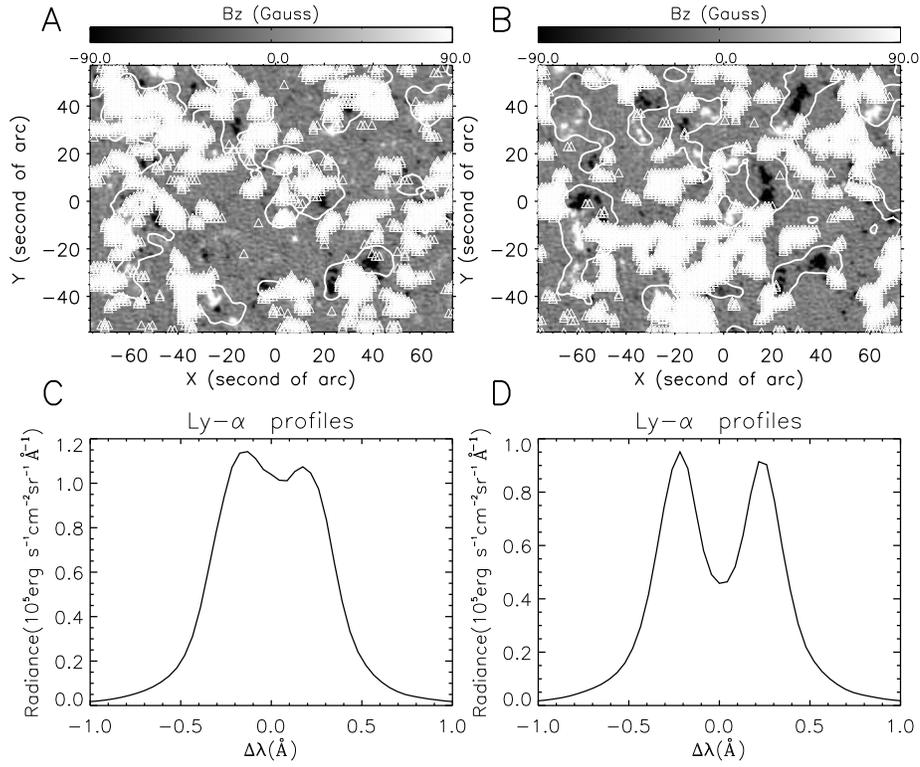}
  \caption {Magnetograms in grey, overlaid with Ly$-\alpha$ brightness contours (white).
  Right: Pixels with the deepest self-revearsal are marked. Left: Pixels with the
  flattest profiles are marked \citep[extreme quartile separated bins, cf. for details,][]{Tian09a}.}
\end{figure}

\section{Conclusion}

We claim that this surprising result is not just a curiosity, but
the manifestation of a fundamental
mass transportation process in the quiet solar atmosphere, which \citet{Foukal78}
introduced as the 'coronal convection'. In the 'blue branch' material is
lifted up in open or quasi-open structures. \citet{McIntosh09}
argue that this may be related to ubiquitious spicular activity. To some part this
material gets trapped in bipolar loops, from where it rains down at both loop
footpoints, which appear bright and redshifted. This 'red branch' is well
documented in \citet{IED08, Curdt08b, Marsch08},
and is also seen in {it Hinode}--EIS data \citep{Tripathy08}.
During phases of higher solar activity a larger portion of quiet
chromospheric network is being replaced by bright plage, which
consequently results in increased irradiance values.

\acknowledgements The SUMER project is financially supported by DLR, CNES, NASA,
and the ESA PRODEX Programme (Swiss contribution). SUMER and MDI are
instruments onboard {\it SOHO}, a mission operated by ESA and NASA.


\begin{thebibliography}{}

\bibitem[Curdt et al.(2008\,a)]{Curdt08a} Curdt, W., Tian, H., Teriaca, L., Sch\"uhle, U. \& Lemaire, P. 2008\,a A\&A 492, L9
\bibitem[Curdt et al.(2008\,b)]{Curdt08b} Curdt, W., Tian, H., Dwivedi, B.N., \& Marsch, E. 2008\,b, A\&A 491, L13
\bibitem[Dammasch et al.(2003)]{IED08} Dammasch, I.E., Curdt,~W., Dwivedi,~B.N., \& Parenti,~S. 2008, Ann.~Geophys. 26, 2955
\bibitem[Foukal(1978)]{Foukal78} Foukal, ~P. 1978, ApJ 223, 1046
\bibitem[Fr\"ohlich(2009)]{Frohlich09} Fr\"ohlich, ~C. 2009, A\&A 501, 27
\bibitem[Fontenla et al.(2002)]{2002} Fontenla,~J.~M., Avrett,~E.~H., \& Loeser,~E. 2002, ApJ 572, 636
\bibitem[Lemaire et al.(2005)]{Lemaire05}Lemaire,~P., Emerich,~C., Vial,~J.-C.,
Curdt,~W., Sch\"uhle,~U., \& Wilhelm,~K. 2005, AdSpR 35, 384
\bibitem[Lemaire(2009)]{Lemaire09} Lemaire,~P. (2009), private communication
\bibitem[Marsch et al.(2008)]{Marsch08} Marsch, ~E., Tian,~H., Sun,~J.,
Curdt,~W., \& Wiegelmann,~T. 2008, ApJ 685, 1262
\bibitem[McIntosh \& de Pontieu(2009)]{McIntosh09} McIntosh,~S. \& de~Pontieu,~B., 2009, ApJ, in press
\bibitem[Kohl et al.(1997)]{Kohl97} Kohl,~J., Noci,~G., Antonucci,~E.,
Tondello,~G., et al. 1997, Sol.~Phys. 175, 613
\bibitem[Sch\"uhle et al.(2000)]{Schuhle00}Sch\"uhle,~U., Hollandt,~J.,
Pauluhn,~A., \& Wilhelm,~K. 2000, in proc. {\it $1^{st}$ Solar and Space Weather
Euroconference}, Santa Cruz de Tenerife, ed. A. Wilson, ESA SP-463, 427
\bibitem[Tian et al.(2009\,a)]{Tian09a} Tian,~H., Curdt,~W., Marsch, E., \& Sch\"uhle, U. 2009\,a, A\&A 504, 239
\bibitem[Tian et al.(2009\,b)]{Tian09b} Tian,~H., Teriaca,~L., Curdt,~W., \& Vial,~J.-C. 2009\,b, ApJ 703, L152
\bibitem[Tripathi et al.(2009)]{Tripathy08} Tripathi,~D., Mason,~H.E., Dwivedi,~B.N., \& del~Zanna,~G. 2009, ApJ 694, 1256
\bibitem[Xia(2003)]{Xia03} Xia, L. 2003, PhD thesis, Georg-August-Universit\"at G\"ottingen

\end{thebibliography}
\end{document}